\documentclass[5p]{elsarticle}

\usepackage{hyperref,amsmath}
\usepackage{amssymb}
\usepackage{color}

\journal{Journal of Magnetism and Magnetic Materials}

\newcommand{\e}{\varepsilon}
\newcommand{\w}{\omega}
\newcommand{\s}{\sigma}

\newcommand{\up}{\uparrow}
\newcommand{\down}{\downarrow}

\begin{document}

\begin{frontmatter}

\title{Majorana-Kondo competition in a cross-shaped double quantum 
dot-topological superconductor system}

\author{Piotr Majek}
\ead{pmajek@amu.edu.pl}

\author{Ireneusz Weymann}

\address{Institute of Spintronics and Quantum Information, Faculty of 
Physics, Adam Mickiewicz University, ul.Uniwersytetu Pozna\'nskiego 2, 61-614 
Pozna{\'n}, Poland}

\begin{abstract}
We examine the transport properties of a 
double quantum dot system coupled to a topological superconducting nanowire 
hosting Majorana quasiparticles at its ends, with the central quantum dot
attached to the left and right leads. We focus on the behavior of the local density of states
and the linear conductance, calculated with the aid of the numerical renormalization  group method,
to describe the influence of the Majorana coupling on the 
low-temperature transport properties induced by the Kondo correlations.
In particular, we show that the presence of Majorana quasiparticles in the 
system affects both the spin-up and spin-down transport channels, affecting the 
energy scales associated with the first-stage and second-stage Kondo temperatures,
respectively, and modifying the low-energy behavior of the system.
\end{abstract}


\end{frontmatter}

\section{Introduction}
Last decades are marked by intensive expansion in almost every area of physics 
research. New topics have started, but several other have been 
revived thanks to the development of theoretical and experimental methods.
Two great examples of the latter are the Majorana fermions and the Kondo effect, 
that have their origin in the 1930s \cite{deHaas1934May, Majorana1937Apr}.
While the nonmonotonic behavior of metals' resistance was explained by Kondo 
and Wilson just a few decades ago \cite{Kondo1964Jul, Wilson1975Oct}, Majorana 
fermions had to wait for proposals from another field of physics, where their 
signatures as exotic quasiparticles can be searched for \cite{Volovik1999Nov, Kitaev2001, Wilczek2009Sep}.
The idea of topological states of matter, which led to possible 
realizations of Majorana quasiparticles, is currently in the centre of 
solid-state physics research \cite{Fu2008Mar, Hasan2010Nov, Qi2011Oct, 
Wang2017Nov, Sato2017May, Prada2020Oct, zhang_NextSteps_2019}. This particular 
field gives promises to use Majorana bound states (MBS),
which are the solid-state equivalents of Majorana fermions, for quantum 
information purposes \cite{Nayak2008Sep, Kitaev2003, Aguado2017}. 
This is due to their robustness against local perturbations.
The potential use in quantum computing lays e.g. in braiding protocols \cite{BraidingReview,Alicea2012Jun}.
One of the first and still very promising setup to realize Majorana bound states 
is the topological superconducting nanowire \cite{Kitaev2001,Alicea2012Jun,Lutchyn2010Aug,Oreg2010Oct,Kim2018May, 
Laubscher2021}. Moreover, several experiments have been done in the pursuit of 
Majorana quasiparticles in such superconductor-semiconductor 
devices \cite{Mourik2012May, Das2012Nov, Deng2012Dec, Albrecht2016Mar, 
Deng2016Dec, jeon_DistinguishingMajorana_2017, Deng2018Aug, Lutchyn2018May}. 
Furthermore, studying the properties of quantum dots interacting with MBS
is also a very fascinating and stimulating theoretical \cite{Leijnse2014Jan, Lee2013Jun, Liu2011Nov, Weymann2017Apr, 
Leijnse2011Oct, Vernek2014Apr, Gong2014Jun, Ricco2019Apr,weymann_majorana-kondo_2020,Majek2021Aug} and 
experimental \cite{Deng2016Dec,Deng2018Aug} field of research.
Recently, similar to MBS, quantum dots were reported to potentially
serve as a qubit \cite{Hendrickx2020Jul,Jirovec2021Aug}.

Taking into account all these similarities, in this paper we focus on the 
intersection of those fields, analyzing the transport properties of a 
multi-impurity system, which is a double quantum dot, and the topological 
superconducting nanowire hosting Majorana bound states at its ends.
In particular, we examine the behavior of the spin-resolved spectral functions
and the linear conductance, which reveal the interplay between the Majorana and Kondo physics.
We show that the coupling to Majorana quasiparicles 
affects both the spin-up and spin-down local density of states, 
changing the characteristic Kondo temperatures.
Moreover, it destroys the second-stage of Kondo screening by
lifting the spectral function at the Fermi energy to 
half of its maximum value when the usual Kondo effect takes place.
These effects are also visible in the behavior of the conductance through the system.

The paper is organized as follows. In Section \ref{sec:model} we describe the 
Hamiltonian of the system. We introduce the method, which is the numerical 
renormalization group procedure. Then, in section \ref{sec:results} we present 
and discuss the numerical results for the spectral functions and conductance.
At the end, in section \ref{sec:conclusions} we summarize the paper.

\section{Model and method}
\label{sec:model}
\begin{figure}[ht!]
	\centering
	\includegraphics[width=0.4\textwidth]{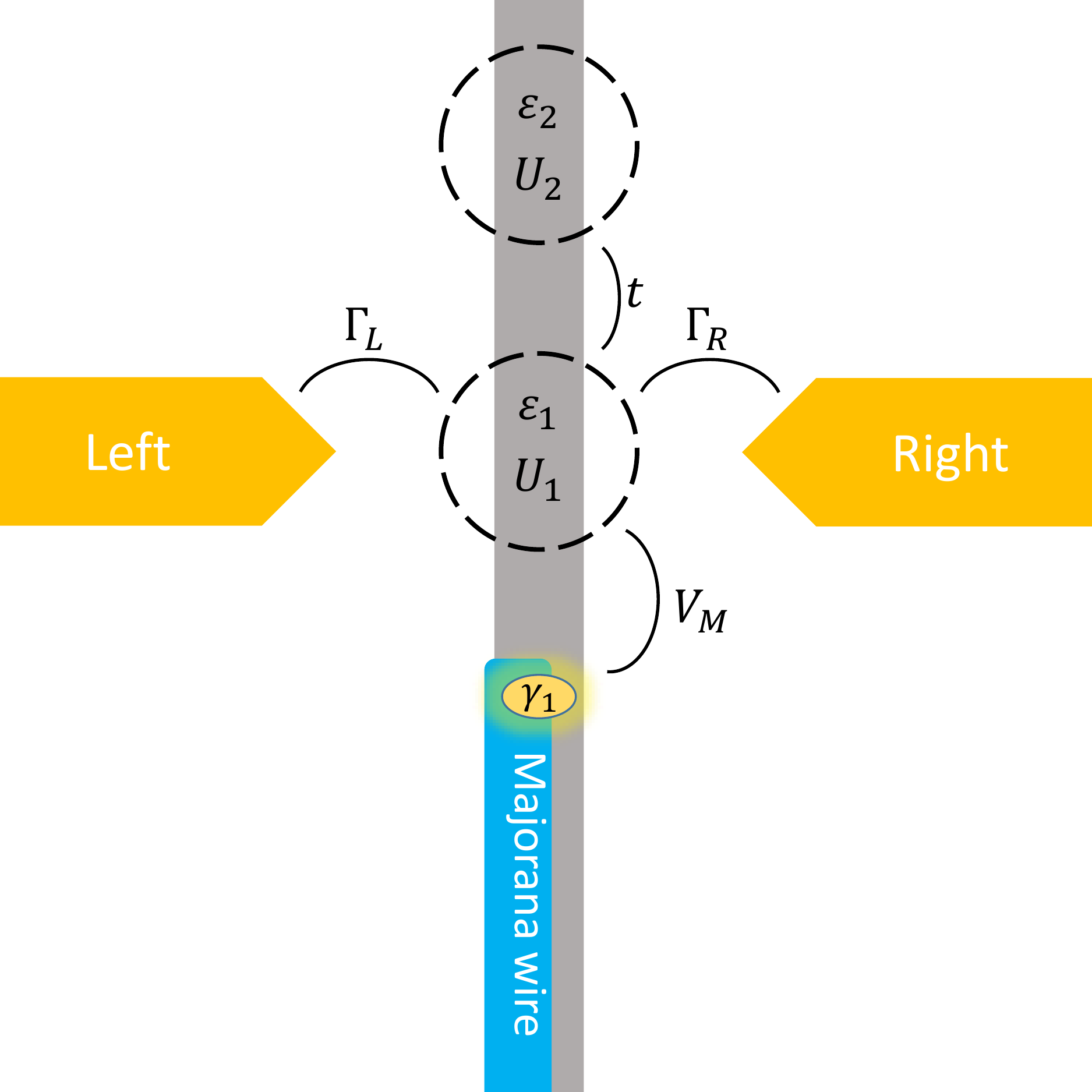}
	\caption{The considered double quantum dot-Majorana wire system. 
		The first quantum dot, described by energy level $\varepsilon_1$ and 
		Coulomb correlations $U_1$, is coupled to two metallic leads by $\Gamma_{r}$ 
		and to the end of topological superconducting nanowire hosting Majorana 
		quasiparticle $\gamma_1$ by matrix elements $V_M$.
		The second quantum dot, parametrized by $\varepsilon_2$ and $U_2$,
		is coupled to the first dot by tunneling matrix elements denoted by $t$.}
	\label{Fig:model}
\end{figure}

The model we are considering in this paper is the Anderson-like double 
quantum dot system, where the first quantum dot is placed in the center of the setup,
coupled to two metallic leads with the strength $\Gamma_{r}$
and to the second quantum dot by tunneling matrix elements $t$.
The first dot is also interacting through matrix elements $V_M$
with $\gamma_1$, which describes the Majorana quasiparticle at the end of topological superconducting nanowire.
The schematic of the model is presented in Fig. \ref{Fig:model}.
The model Hamiltonian consists of four parts:
\begin{equation}\label{eq:Hamiltonian}
	H = H_{L} + H_{T} + H_{DQD} + H_{M}
\end{equation}
with
\begin{equation}\label{eq:Hleads}
	H_{L} = \sum_{r=L,R}\sum_{\mathbf{k}\sigma}
	\e_{r\mathbf{k}} c^\dag_{r\mathbf{k}\sigma} c_{r\mathbf{k}\sigma}
\end{equation}
describing the left ($r$ = L) and right ($r$ = R) lead as a reservoir of 
non-interacting quasiparticles, where 
$c_{r\textbf{k}\sigma}^\dagger$ is the creation operator of an electron in the 
lead $r$ with momentum $\textbf{k}$, spin $\sigma$ and energy 
$\varepsilon_{r\textbf{k}}$. The next part describes the tunneling between  
leads and the first quantum dot
\begin{equation}
	H_T = \sum_{r=L,R}\sum_{\mathbf{k}\sigma} v_{r} \left(d^\dag_{1\s}
	c_{r\mathbf{k}\sigma} + c^\dag_{r\mathbf{k}\sigma} d_{1\s} \right),
\end{equation}
with $d_{1\sigma}^\dagger$ denoting the creation operator of an electron with 
spin $\sigma$ on the first dot.
The coupling between the dot and the lead is defined by 
tunneling rate $\Gamma_r = \pi \rho v_r^2$, with $\rho$ being the density 
of states of the lead $r$. The double quantum dot is described by
\begin{multline}
	H_{DQD} = \sum_{j=1,2} \sum_{\s} \e_j d_{j\s}^\dag d_{j\s} \\ 
	+ \sum_{j=1,2} U_j d_{j\uparrow}^\dag d_{j\uparrow} 
	d_{j\downarrow}^\dag d_{j\downarrow} \\
	+ \sum_\s t (d_{1\s}^\dagger d_{2\s} +  d_{2\s}^\dag d_{1\s}).
\end{multline}
Here, $d_{j\sigma}^\dagger$ creates an electron with spin $\sigma$ on the 
$j$-th quantum dot with energy $\varepsilon_j$. The Coulomb interaction between 
electrons on the $j$-th dot is denoted by $U_j$, and hopping between the dots is 
described by the hopping matrix elements $t$. The last part couples the first dot and 
Majorana wire
\begin{equation}
	H_M =  \sqrt{2} V_M (d^\dag_{1\downarrow} \gamma_1 + \gamma_1 
	d_{1\downarrow})
\end{equation}
where $\gamma_1$ is the Majorana bound state at the end of the nanowire and 
$V_M$ denotes the relevant tunneling matrix elements
\cite{Lee2013Jun,Liu2011Nov,Weymann2017Apr,weymann_majorana-kondo_2020,Flensberg2010Nov,Weymann2017Jan}.

The Majorana quasiparticle can be expressed with the usual fermion operator $f$ as follows 
$\gamma_1 = (f^\dagger + f)/\sqrt{2}$. For the further calculations we rewrite 
this part of the Hamltonian with the fermionic operators, obtaining
\begin{equation}
	H_M = V_M (d_{1 \down}^\dagger - d_{1 \down}) (f^\dagger + f).
\end{equation}

Topological superconducting nanowire hosts a pair of Majorana bound states,
however, in our calculations we consider the long nanowire case,
where the two Majorana mode wavefunctions do not overlap.
Consequently, the second Majorana mode $\gamma_2$ do not show up in our model.
We also mention the assumption, that only the spin-down electrons are 
coupled with the Majorana wire \cite{Lee2013Jun}.

In this paper we are examining the spectral functions of the considered system,
which give insight into the system's transport behavior.
Because in the studied setup the first dot is attached to the leads, 
we focus on the behavior of the spin-$\sigma$ spectral function of the first 
quantum dot, which is defined as $A_\sigma(\omega) = - \frac{1}{\pi}{\rm Im} 
G_\sigma^R(\omega)$. Here, $G_\sigma^R(\omega)$ is the Fourier transform of 
the retarded Green's function $G_\sigma^R (t) = - i \Theta(t) \langle \{ 
d_{1\sigma} (t), d_{1\sigma}^\dagger (0) \} \rangle$.
The spectral function is directly related to the linear conductance through the system,
which is given by, $G=G_\up + G_\down$, with
\begin{equation}
G_\sigma = \frac{e^2}{h} \frac{4\Gamma_L \Gamma_R}{\Gamma_L + \Gamma_R} \int d\omega \pi A_\sigma(\omega)[-f'(\omega)],
\end{equation}
where $f'(\omega)$ denotes the derivative of the Fermi function.
Due to strong electron correlations in the model, we are using the numerical renormalization group (NRG) 
procedure introduced by Wilson \cite{Wilson1975Oct},
which allows for obtaining reliable results for the system's behavior
\cite{Bulla2008Apr, NRG_code}.

\section{Numerical results and discussion}
\label{sec:results}

In this section we present the numerical results obtained with the aid of the NRG. 
The system was set up with the following parameters:
the Coulomb interaction $U_1= U_2 \equiv U = 0.2D$,
where $D$ is the half bandwidth used as energy unit,
quantum dots' energy levels tuned to the particle-hole symmetry $\varepsilon_1 
= \varepsilon_2 = - U/2$, $\Gamma = \Gamma_L+\Gamma_R = 0.1U$,
and $t$, $V_M$ as described in the figures. 
The NRG parameters we have used were: the discretization parameter $\Lambda = 
2$ and the number of states kept during $60$ NRG iterations was $3072$.

\begin{figure}[t!]
	\center
	\includegraphics[width=0.49\textwidth]{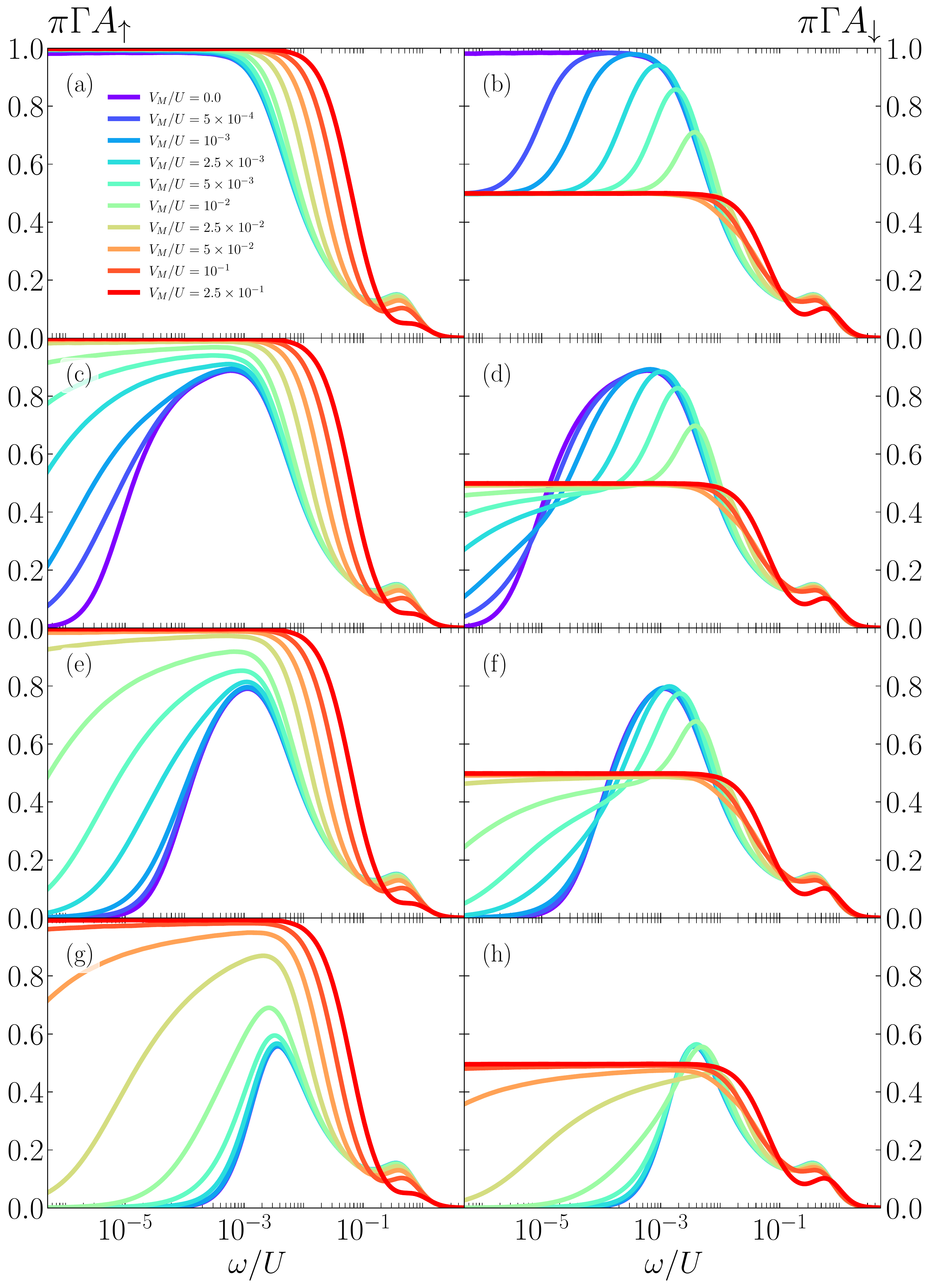}
	\caption{The spin-up (left column) and spin-down (right column) spectral 
	functions calculated for (a, b) $t/U = 0.01$, (c, d) $t/U = 0.02$, (e, f) 
	$t/U = 0.025$, (g, h) $t/U = 0.0375$ and multiple values of $V_M$, as indicated in the legend.
	The other parameters are: $U_1=U_2\equiv U=0.2$, $\e_1=\e_2=-U/2$,
	and $\Gamma = 0.1 U$.} 
	\label{Fig:spectral_functions_t_vs_VM}
\end{figure}

\begin{figure}[t!]
	\center
	\includegraphics[width=0.49\textwidth]{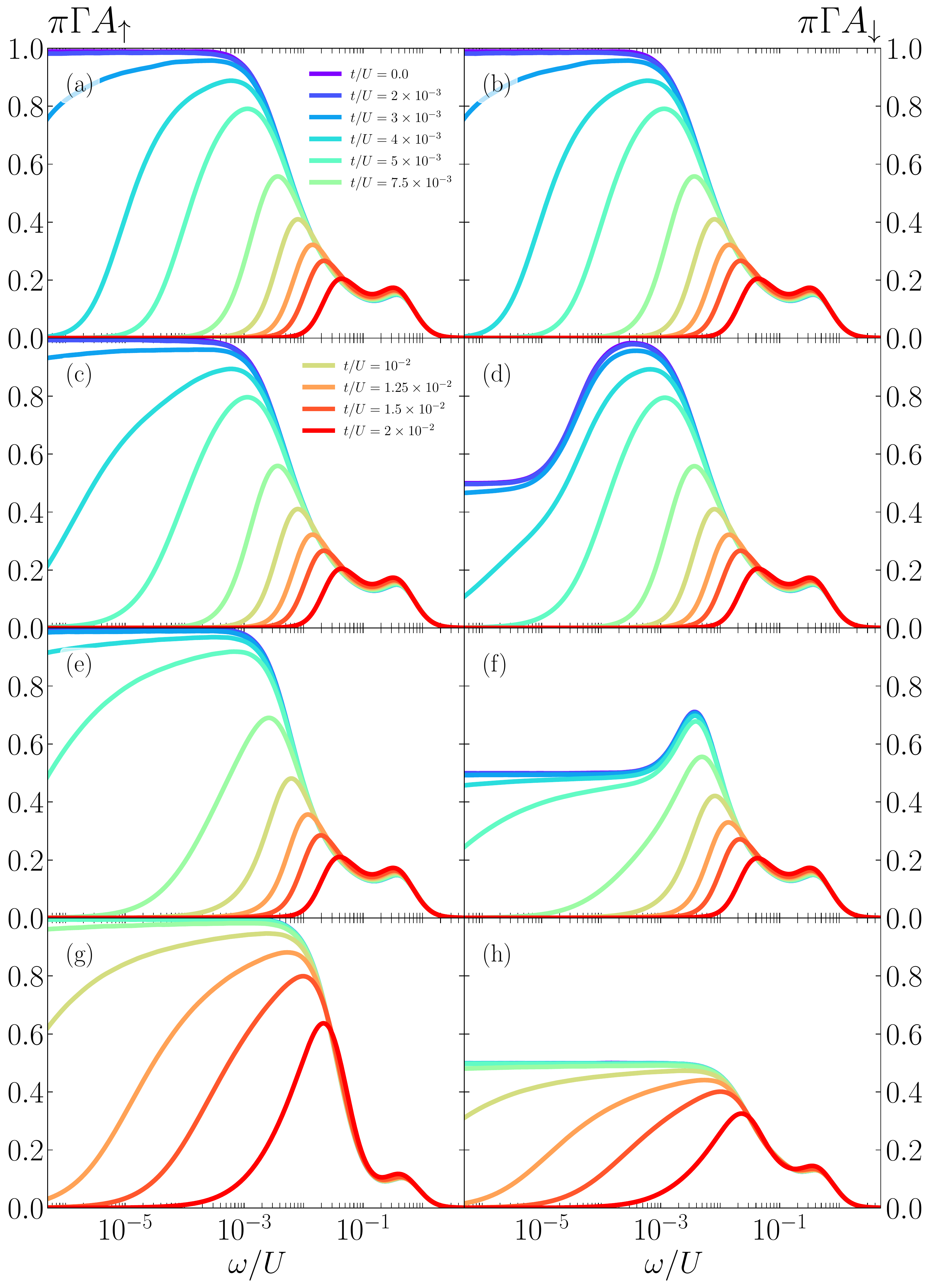}
	\caption{The spin-up (left column) and spin-down (right column) spectral 
		functions calculated for (a, b) $V_M/U = 0.0$, (c, d) $V_M/U = 0.001$, 
		(e, f) 
		$V_M/U = 0.01$, (g, h) $V_M/U = 0.1$ and different values of $t$,
		as indicated. The other parameters are the same as in Fig.~\ref{Fig:spectral_functions_t_vs_VM}.}
	\label{Fig:spectral_functions_VM_vs_t}
\end{figure}

In Fig. \ref{Fig:spectral_functions_t_vs_VM} we present the spectral functions 
calculated for fixed value of hopping between the quantum dots and different values of coupling to 
the Majorana wire. The first row shows the case when a weak hopping between the dots is assumed.
Then, the spin-up spectral function shows the usual single-impurity Kondo effect, 
where the Kondo temperature can be estimated from \cite{Haldane1978Feb}
\begin{equation}\label{eq:TK}
	T_K \approx \sqrt{\frac{\Gamma U}{2}} \exp{\left[ \frac{\pi}{2} \frac{\e_1 
			(\e_1 + 
			U)}{\Gamma U} \right]}
\end{equation}
and approximately corresponds to the half width at half maximum of the spectral function.
The significant influence of the presence of MBS in the system in the weak hopping regime 
is the enhancement of $T_K$ \cite{Weymann2017Apr}. Much different picture can be 
observed in the behavior of the spin-down component of the spectral function, see 
Fig. \ref{Fig:spectral_functions_t_vs_VM}(b).
It can be seen that finite coupling to Majorana wire suppresses
the low energy spectral function to a finite value of $1/2\pi\Gamma$,
whereas a local maximum of height $1/\pi\Gamma$
remains in certain energy range.
When the coupling $V_M$ becomes considerable
($V_M \gtrsim t$), the spectral function becomes fully suppressed to its half value.
On the other hand, when the hopping between the dots becomes stronger,
the second stage of the Kondo screening occurs in the system, which
is revealed through the suppression of the spectral function to zero,
see the curve for $V_M=0$ in Figs. \ref{Fig:spectral_functions_t_vs_VM}(c,d).
The second-stage Kondo temperature can be defined as 
\cite{Cornaglia2005Feb}
\begin{equation}\label{eq:T*}
	T^* \approx \alpha T_K \exp{\left( -\beta T_K/J_{\rm{eff}} \right)}
\end{equation}
with dimensionless constants $\alpha, \beta$ of the order of unity, and 
$J_{\rm{eff}} = 4t^2/U$.
When the coupling to topological superconductor is turned on, $T_K$
behaves in a similar way as for the weak hopping regime,
being increased as $V_M$ is getting stronger.
Interestingly, with increasing $V_M$, the second stage of screening 
becomes suppressed, such that $A_\up(0) = 1/\pi \Gamma$.
This behavior changes for the spin-down channel,
which reveals the half-fermionic character of the coupling to MBS.
From the usual two-stage process, the spectral function $A_\down(\omega)$ is being lifted in 
the low-energy scales, and suppressed for higher $\w$, eventually reaching the half of its maximum value.
The above described scenario can also be observed 
when the hopping between the dots is increased,
however, larger values of $V_M$ are needed to suppress
the second stage of the Kondo screening.

In Fig. \ref{Fig:spectral_functions_VM_vs_t} we present the same spectral 
functions, but now plotted for fixed values of coupling to Majorana quasiparticle, while
varying the value of hopping between the quantum dots.
The first row shows the case of the bare double quantum dot system,
where one can see that the increase of $t$ causes 
the low-temperature suppression of the spectral function, which results from the second 
stage of the Kondo effect. Finite interaction between the Majorana wire and 
DQD system shows the competitive character of these two couplings, where the 
low-energy value of spectral function is being restored.
This behavior depends on the ratio of the two couplings, $V_M$ and $t$,
what can be seen particularly for the largest values of either $V_M$ or $t$.
When the hopping $t$ does not reach the order of magnitude of $V_M$, the Majorana 
coupling affects only the spectral function behavior at low energies.
However, higher values of coupling to the MBS may give rise to the case
when the influence of Majorana quasiparticle in the 
system is being stronger than the Kondo correlations, destroying the 
behavior related with the second-stage Kondo effect.
These properties again differ between the spin channels,
where one can see that not only the second-stage Kondo temperature is being affected,
but also the spin-down spectral function becomes half-suppressed
as the Majorana coupling is strong enough to 
exceed the exchange interaction between the quantum dots. It is also important 
to note that, similarly as in the case of 
Fig. \ref{Fig:spectral_functions_t_vs_VM},
with an increase of $V_M$, the first-stage Kondo temperature $T_K$ becomes enhanced.
This can be especially seen as a shifting of the steep increase (due to Kondo correlations)
of the spin-up spectral function to larger energies, which is visible in the left column
of Fig. \ref{Fig:spectral_functions_VM_vs_t}.

\begin{figure}[t!]
	\center
	\includegraphics[width=0.49\textwidth]{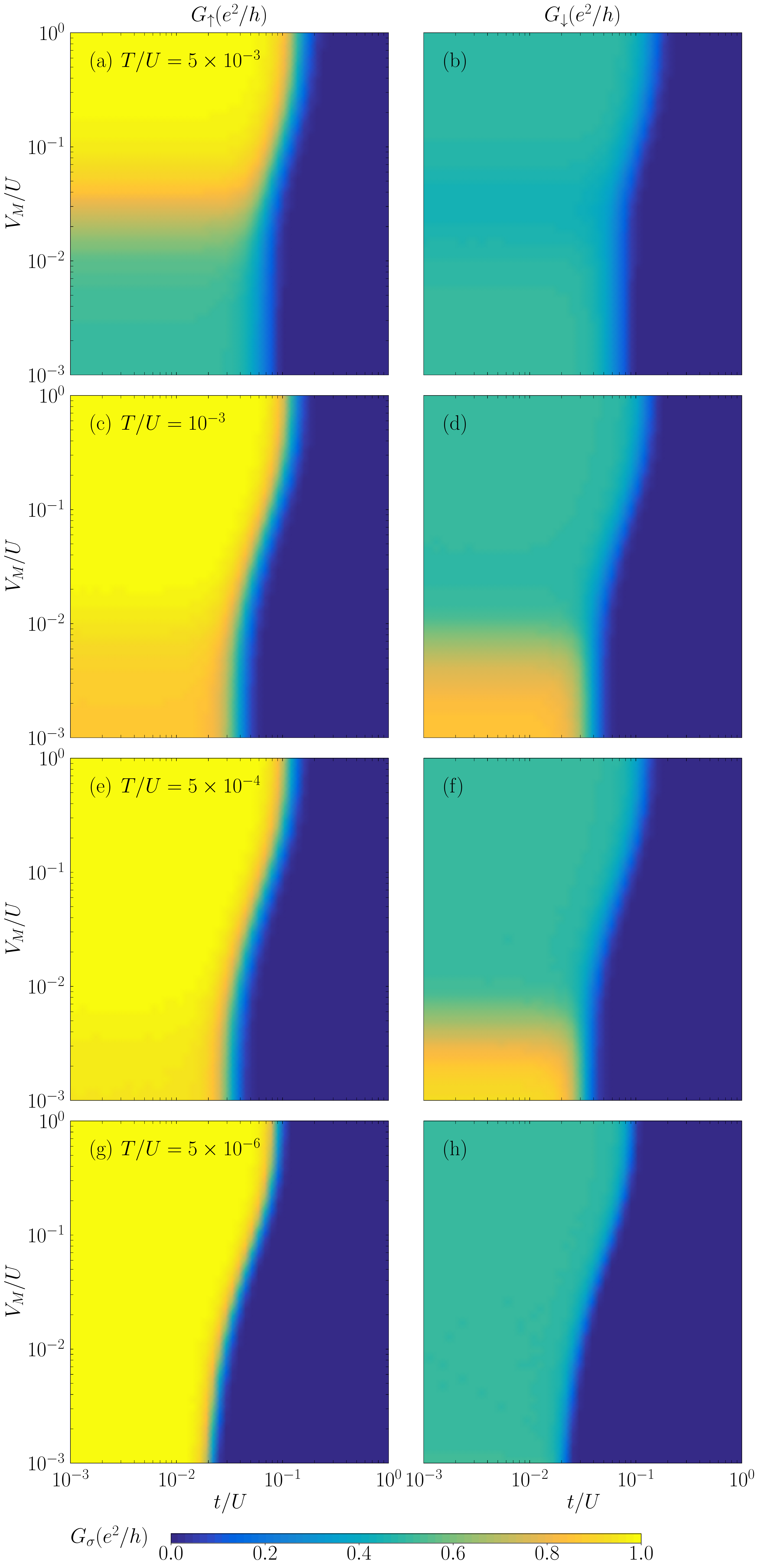}
	\caption{The linear conductance calculated as a function
		of the hopping between the dots $t$ and the coupling
		to Majorana wire $V_M$ for different values of temperature, as indicated.
		The left column presents $G_\uparrow$, while
		the right column shows $G_\downarrow$.
		The other parameters are the same
		as in Fig.~\ref{Fig:spectral_functions_t_vs_VM}.}
	\label{Fig:G}
\end{figure}
Finally, in Fig.~\ref{Fig:G} we present the spin-resolved linear conductance
calculated as a function of the coupling to Majorana wire $V_M$
and the hopping between the dots $t$ for different temperatures, as indicated in the figure.
Consider first the case of lowest temperature presented in the bottom row of Fig.~\ref{Fig:G}.
When the hopping between the dots
is relatively weak, $G_\up = e^2/h$ due to the Kondo effect, while 
$G_\down = e^2/2h$ due to the quantum interference with the Majorana wire.
Increasing $t$ results in an enhancement of the second-stage Kondo temperature $T^*$,
and once $T^*\gtrsim T$, the second-stage Kondo effect comes into play.
As a consequence, the conductance in both spin channels becomes suppressed.
Note also that the value of $t$, at which the conductance drops,
increases with raising the coupling to the Majorana wire,
which clearly indicates that $V_M$ affects the Kondo behavior in the system.
When the temperature increases, one observes a change in the conductance
for low values of both $t$ and $V_M$, see the case of $T/U=5\times 10^{-4}$
and $T/U=10^{-3}$ in Fig.~\ref{Fig:G}.
In the spin-up channel the conductance becomes suppressed, while in the spin-down channel
it gets enhanced. This is due to the fact that now thermal fluctuations
smear out the behavior, as described above,
resulting from the quantum interference with Majorana wire.
Interestingly, in the case when $T/U = 5\times 10^{-3}$,
shown in the first row of Fig.~\ref{Fig:G},
the behavior of conductance is even more changed.
Because in this situation $T\approx T_K$, $G_{\down}$ hardly depends on $V_M$.
This can be understood by analyzing the spectral function behavior
presented in the right column of Fig.~\ref{Fig:spectral_functions_t_vs_VM}
for $\omega \approx T$. At this energy, $A_\down(\omega\approx T)\approx 1/2\pi\Gamma$,
and it hardly depends on $V_M$, which results in $G_\down\approx e^2/2h$,
as long as the hopping between the dots is relatively weak.
On the other hand, the spin-up conductance
exhibits a different behavior. Clearly, we see an increase 
of $G_\up$ with raising $V_M$ for low values of $t$.
Again, this can be understood by inspecting the 
dependence of the spin-up spectral function on $V_M$
at the energy of the order of assumed temperature, 
see the left column of Fig.~\ref{Fig:spectral_functions_t_vs_VM}.

\section{Conclusions}
\label{sec:conclusions}
In this paper we have analyzed the behavior of the Kondo correlated double 
quantum dot system coupled to the superconducting topological nanowire hosting 
Majorana bound state in the cross-shaped geometry, where one quantum 
dot is in the center of the system, being coupled to the leads.
We have focused on the transport properties of the system in the Kondo regime,
where the two-stage Kondo effect is present. We have presented and discussed the behavior of the spectral 
functions and the conductance, where the aforementioned effects can be found. The results were 
obtained with the aid of numerical renormalization group method, which allowed 
us to analyze the transport properties in an accurate manner.

In our work we have uncovered the most interesting aspects of coupling the DQD 
system to the MBS. A fierce competition between the Kondo and Majorana physics 
is present not only in the spin-down electron channel, which is directly 
coupled to the nanowire, but also the spin-up channel reveals a strong 
influence of the Majorana coupling in the system. Not only the restoring 
of the low-energy spectral behavior has been shown, but also the 
half-fermionic character of the MBS, which demonstrates itself when the 
spin-down spectral function is restored only to the half of its total value.
Moreover, the evolution of the characteristic fractional value of conductance
with the hopping between the dots, coupling to Majorana wire and the temperature was analyzed.
The presented numerical analysis reveals an interesting
competitive character of interactions driving the Kondo and Majorana physics.

\section*{Acknowledgemenets}
This work was supported by the National Science Centre in Poland through the 
Project No. 2018/29/B/ST3/00937. The computing time at the Pozna\'n 
Supercomputing and Networking Center is acknowledged.


\end{document}